\documentclass[a4,aps,prd,superscriptaddress,floatfix,nofootinbib,showpacs]
{revtex4}

\usepackage{color}
\usepackage[dvips]{graphicx}

\newcommand{\lsim}{\mathrel{\mathop{\kern 0pt \rlap
  {\raise.2ex\hbox{$<$}}}
  \lower.9ex\hbox{\kern-.190em $\sim$}}}
\newcommand{\gsim}{\mathrel{\mathop{\kern 0pt \rlap
  {\raise.2ex\hbox{$>$}}}
  \lower.9ex\hbox{\kern-.190em $\sim$}}}

\newcommand{\be}{\begin{equation}}
\newcommand{\ee}{\end{equation}}
\newcommand{\beqarr}{\begin{eqnarray}}
\newcommand{\eeqarr}{\end{eqnarray}}

\begin{document}


\title{Lower Bound on the  Neutralino Mass from New Data on CMB \\
and Implications for Relic Neutralinos}


  
%
\author{A. Bottino}
\affiliation{Dipartimento di Fisica Teorica, Universit\`a di Torino \\
Istituto Nazionale di Fisica Nucleare, Sezione di Torino \\
via P. Giuria 1, I--10125 Torino, Italy}

\author{F. Donato}
\affiliation{Dipartimento di Fisica Teorica, Universit\`a di Torino \\
Istituto Nazionale di Fisica Nucleare, Sezione di Torino \\
via P. Giuria 1, I--10125 Torino, Italy}

\author{N. Fornengo}
\affiliation{Dipartimento di Fisica Teorica, Universit\`a di Torino \\
Istituto Nazionale di Fisica Nucleare, Sezione di Torino \\
via P. Giuria 1, I--10125 Torino, Italy}

\author{S. Scopel} 
\email{bottino@to.infn.it, donato@to.infn.it, fornengo@to.infn.it, scopel@to.infn.it}
\homepage{http://www.to.infn.it/astropart}
\affiliation{Dipartimento di Fisica Teorica, Universit\`a di Torino \\
Istituto Nazionale di Fisica Nucleare, Sezione di Torino \\
via P. Giuria 1, I--10125 Torino, Italy}

\date{\today}

\begin{abstract} \vspace{1cm}  
In the frame of an effective MSSM model without gaugino-mass
unification at a grand unification scale, we set a lower bound
on the neutralino mass based on the new WMAP data on
$\Omega_{CDM}$ (R-parity conservation is assumed). 
Our lower bound, $m_{\chi} \gsim 6$ GeV, leaves much
room for relic neutralinos significantly lighter than those commonly
considered ($m_{\chi} \gsim 50$ GeV).  We prove that these light
neutralinos may produce measurable effects in WIMP direct detection
experiments of low energy threshold and large exposure.
\end{abstract}


\maketitle

\section{Introduction}
\label{sec:intro}

In a previous paper \cite{light}, we called the attention to relic
neutralinos of light masses, {\it i. e.} with a mass $m_{\chi} \lsim
45$ GeV.  Actually, most analyses on cosmological neutralinos employ a
lower bound $m_{\chi} \gsim 50$ GeV, {\it i. e.} a bound which rests
on the assumption that gaugino masses are unified at the 
grand unification (GUT) scale
$M_{GUT} \sim 10^{16}$ GeV. However, this unification assumption might
not be justified, as already discussed some time ago
\cite{eent,d}. Furthermore, recent analyses of string models (see, for
instance, Ref. \cite{munoz}) indicate that the initial scale for the running
of the susy parameters by the renormalization group equations may be
much lower than the standard GUT-scale, with the implication that at
the electroweak (EW) scale the gaugino masses may be quite different
from what expected in a standard SUGRA scheme with GUT-unification
assumptions.  Thus, for instance, the U(1) and SU(2) gaugino masses,
$M_1$ and $M_2$, may not be related by the standard formula $M_1
\simeq \frac{1}{2} M_2$ at the EW scale.

It is then quite natural to discuss supersymmetric models where $M_1$
and $M_2$ are considered as independent parameters. Previous papers
where schemes of this type are considered include the ones in
Refs. \cite{eent,d,dt,gr,mny,go,bbd,cn,ads,cerdeno,bk,bbb,bno,bhn,boudjema,hooper}.
In Ref. \cite{light} we evaluated the neutralino relic abundance
$\Omega_{\chi} h^2$ and the neutralino-nucleon scalar cross-section
$\sigma_{\rm scalar}^{(\rm nucleon)}$ in an effective Minimal
Supersymmetric extension of the Standard Model (effMSSM) where the
gaugino-mass unification at GUT scale is not assumed. The analysis was
performed in a scenario where the ratio $R \equiv M_1/M_2$ is smaller
than the GUT-unification value $R_{GUT} \simeq 0.5$, thus allowing the
neutralino mass $m_{\chi}$ to be lighter than the commonly-used lower
bound of about 50 GeV.  We showed that, in the derivation of the lower
limit on $m_{\chi}$, the upper bound on the relic abundance for cold
dark matter (CDM), $\Omega_{CDM} h^2$, plays a crucial role. A lower limit of 
 $m_{\chi} \gsim $ 5 GeV was derived in Ref. \cite{light}, by
employing the upper bound $\Omega_{CDM} h^2 \lsim 0.3$. It was also
shown how this upper bound on $\Omega_{CDM} h^2$ is instrumental in
providing sizeable values for the neutralino-nucleon scalar
cross-section at small $m_{\chi}$.

New data on the cosmic microwave background (CMB)
\cite{wmap,cbi,acbar}, also used in combination with other
cosmological observations, are progressively narrowing down the ranges
of the relic abundances for matter ($\Omega_m h^2$) and for some of
its constituents: neutrinos ($\Omega_{\nu} h^2$) and baryons
($\Omega_b h^2$). As a consequence also the range of $\Omega_{CDM}
h^2$ is reaching a unprecedented level of accuracy.  The importance of
an improvement in the determination of $\Omega_{CDM} h^2$ for any cold
relic particle is twofold. The upper bound $(\Omega_{CDM} h^2)_{max}$
obviously establishes a strict upper limit for any specific cold
species.  On the other side, the lower bound $(\Omega_{CDM}
h^2)_{min}$ fixes the value of the average abundance below which the
halo density of a specific cold constituent has to be rescaled as
compared to the total CDM halo density.  For the determination of the
rescaling factor $\xi \equiv \rho_{\chi}/\rho_0$ (where $\rho_{\chi}$
and $\rho_0$ are the local neutralino density and the total local dark
matter density, respectively), we use the standard rescaling recipe
$\xi = {\rm min}(1,\Omega_\chi h^2/(\Omega_{CDM} h^2)_{\rm min})$.  In
Ref. \cite{wmap} ranges for $\Omega_m h^2$ and $\Omega_b h^2$ are
derived, by employing CMB data of Refs. \cite{wmap,cbi,acbar}, 2dFGRS
measurements \cite{2dFGRS} and Lyman $\alpha$ forest data
\cite{forest}. From the values quoted in Ref. \cite{wmap} for
$\Omega_m h^2$ and $\Omega_b h^2$ and allowing for a $2-{\sigma}$
range in $\Omega_{CDM} h^2$, one obtains: $(\Omega_{CDM} h^2)_{min} =
0.095$ and $(\Omega_{CDM} h^2)_{max} = 0.131$.  These are the values
we will use here. However, one should cautiously still be open to some
possible changes in these values as new cosmological observational
data will accumulate in the future.

In the present paper we analyse the supersymmetric scenario of
Ref. \cite{light} in light of these new determinations of
$\Omega_{CDM} h^2$.  We first derive approximate analytic expressions
which directly relate the lower bound on $m_{\chi}$ to the upper
limit on $\Omega_{CDM} h^2$ (R-parity conservation is assumed). 
This requires the evaluation of
$\Omega_\chi h^2$, which is in itself a very complicated function of
the supersymmetric parameters. In order to extract the leading
analytic terms at small values of $m_{\chi}$, we are guided by a
detailed numerical analysis of the supersymmetric parameter
space. This will allow us to present analytic expressions for the
lower bounds on $m_{\chi}$ which display the link of these bounds to
the relevant particle-physics and cosmological constraints, such as the 
lower bounds on sfermions and Higgs-bosons masses and the upper limit on 
$\Omega_{CDM} h^2$.  Furthermore, on the basis of these results, we
show that light neutralinos, with masses $m_{\chi} \lsim $ 45 GeV, may
actually be probed by WIMP direct detection experiments with high
sensitivities and low energy thresholds.

\section{Lower bound on $m_{\chi}$} 

The supersymmetric scheme adopted here is the same as the one
described in Ref. \cite{light}: an effective MSSM scheme (effMSSM) at
the electroweak scale, with the following independent parameters:
$M_2, \mu, \tan\beta, m_A, m_{\tilde q}, m_{\tilde l}, A$ and $R
\equiv M_1/M_2$. Notations are as follows: $\mu$ is the Higgs mixing mass
parameter, $\tan\beta$ the ratio of the two Higgs v.e.v.'s, $m_A$ the
mass of the CP-odd neutral Higgs boson, $m_{\tilde q}$ is a quark
soft--mass common to all squarks, $m_{\tilde l}$ is a slepton
soft--mass common to all sleptons, $A$ is a common dimensionless
trilinear parameter for the third family, $A_{\tilde b} = A_{\tilde t}
\equiv A m_{\tilde q}$ and $A_{\tilde \tau} \equiv A m_{\tilde l}$
(the trilinear parameters for the other families being set equal to
zero).

The neutralino is defined as the lowest--mass linear superposition of
bino $\tilde B$, wino $\tilde W^{(3)}$ and of the two higgsino states
$\tilde H_1^{\circ}$, $\tilde H_2^{\circ}$:
\begin{equation}
\chi \equiv a_1 \tilde B +
a_2 \tilde W^{(3)} + a_3 \tilde H_1^{\circ} + a_4 \tilde H_2^{\circ}.
\end{equation}

Since we are here interested in light neutralinos, we consider values
of $R$ lower than its GUT value: $R_{GUT} \simeq 0.5$; for
definiteness, we take $R$ in the range: 0.01 - 0.5.

We first outline the procedure for deriving analytical bounds on
$m_{\chi}$ from the cosmological upper limit on $\Omega_{CDM} h^2$.
As mentioned in the previous section, the identification of the
leading analytic contributions is guided by numerical analysis. This
is based on a scanning of the supersymmetric parameter space, with the
following ranges of the MSSM parameters: $1 \leq \tan \beta \leq 50$,
$100\, {\rm GeV }\leq |\mu|, M_2, m_{\tilde q}, m_{\tilde l} \leq
1000\, {\rm GeV }$, ${\rm sign}(\mu)=-1,1$, $90\, {\rm GeV }\leq m_A
\leq 1000\, {\rm GeV }$, $-3 \leq A \leq 3$.  The following
experimental constraints are imposed: accelerators data on
supersymmetric and Higgs boson searches (CERN $e^+ e^-$ collider LEP2
\cite{LEPb} and Collider Detector CDF at Fermilab \cite{cdf});
measurements of the $b \rightarrow s + \gamma$ decay \cite{bsgamma};
measurements of the muon anomalous magnetic moment $a_\mu \equiv
(g_{\mu} - 2)/2$ \cite{amm} (the range $-160 \leq \Delta a_{\mu} \cdot
10^{11} \leq 680 $ is used here for the deviation of the current
experimental world average from the theoretical evaluation within the
Standard Model; for the derivation see Ref. \cite{light}).

The neutralino relic abundance is given by 

\begin{equation}
\Omega_{\chi} h^2 = \frac{x_f}{{g_{\star}(x_f)}^{1/2}} \frac{3.3 \cdot
10^{-38} \; {\rm cm}^2}{\widetilde{<\sigma_{ann} v>}},
\label{omega} 
\end{equation}

\noindent
where $\widetilde{<\sigma_{ann} v>} \equiv x_f {\langle \sigma_{\rm
ann} \; v\rangle_{\rm int}}$, ${\langle \sigma_{\rm ann} \;
v\rangle_{\rm int}}$ being the integral from the present temperature up to
the freeze-out temperature $T_f$ of the thermally averaged product of
the annihilation cross-section times the relative velocity of a pair
of neutralinos, $x_f$ is defined as $x_f \equiv \frac{m_{\chi}}{T_f}$ and
${g_{\star}(x_f)}$ denotes the relativistic degrees of freedom of the 
thermodynamic
bath at $x_f$.  For $\widetilde{\langle \sigma_{\rm ann} \; v\rangle}$
we will use the standard expansion in S and P waves:
$\widetilde{\langle \sigma_{\rm ann} \; v\rangle}\simeq \tilde{a} +
\frac{1}{2 x_f} \tilde{b}$.

A lower bound on $m_{\chi}$ is now derived from Eq. (\ref{omega}), by
requiring that

\begin{equation}
\Omega_{\chi} h^2 \leq (\Omega_{CDM} h^2)_{max}.
\label{r}
\end{equation}

We first work out approximate analytic expressions for
$\widetilde{<\sigma_{ann} v>}$ in the regime of small $m_{\chi}$
($m_{\chi} \lsim $ 45 GeV).  By diagonalizing the usual neutralino
mass matrix in the approximation $M_1 << M_2, \mu$, it turns out that
light neutralinos have a dominant bino component; a deviation from a
pure bino composition is mainly due to a mixture with $\tilde
H_1^{\circ}$, {\it i. e.}  $|a_1| >> |a_3| >> |a_1|, |a_4|$. For the
ratio $|a_3|/|a_1|$ one finds

\begin{equation}
\frac{|a_3|}{|a_1|} \simeq \sin \theta_{W} \; \sin \beta \;
\frac{m_Z}{\mu} \lsim 0.42 \; \sin \beta,
\label{ratio}
\end{equation}
\noindent
where in the last step we have taken into account the experimental
lower bound $\mu \gsim $ 100 GeV.

The dominant terms in $\widetilde{\langle \sigma_{\rm ann} \;
 v\rangle_{\rm int}}$ are the contributions due to Higgs-exchange in
 the $s$ channel and sfermion-exchange in the $t,u$ channels of the
 annihilation process $\chi + \chi \rightarrow \bar{f}+f$
 (interference terms are neglected). We retain only the leading terms
 in each contribution. Thus, for the Higgs--exchange contribution,
 dominated by the S-wave annihilation into down-type fermions, we
 have, for any final state $\bar{f}-f$,

\begin{equation}
{\widetilde{\langle \sigma_{\rm ann} \; v\rangle}_f}^{Higgs}\simeq
{\tilde{a}_f}^{Higgs} \simeq \frac{2 \pi \alpha_{e.m.}^2 c_f}{\sin^2 \theta_W
\; \cos^2 \theta_W} a_1^2 a_3^2 \tan^2 \beta ~ (1 + \epsilon_f)^2 ~
\frac{\bar{m}_f^2}{m_W^2}
\frac{m_{\chi}^2~[1-m_f^2/m_\chi^2]^{1/2}}{[(2m_{\chi})^2 - m_A^2]^2},
\label{ah}
\end{equation}

\noindent
where $c_f$ is a standard color factor ($c_f = 3$ for quarks, $c_f =
 1$ for leptons), $\bar{m}_f$ is the fermion running mass evaluated
 at the energy scale $2 m_{\chi}$ and $m_f$ is the fermion pole mass.
$\epsilon_f$ is a quantity which
 enters in the relationship between the down--type fermion running
 mass and the corresponding Yukawa coupling (see Ref. \cite{higgs} and
 references quoted therein); in the following evaluations,
 $\epsilon_f$ is negligible, except for the bottom quark, where
 $\epsilon_b \simeq 0.2$.  One easily verifies that when $m_{\chi} <
 m_b$, ${\widetilde{\langle \sigma_{\rm ann} \; v\rangle}_f}^{Higgs} $
 entails a relic abundance exceeding the cosmological bound.

Notice that $\widetilde{<\sigma_{ann} v>}$ turns out to be an
increasing function of $m_{\chi}$. Then, to obtain a conservative
lower bound on $m_{\chi}$ from the condition of Eq. (\ref{r}), we have
first to evaluate an $(\Omega_{\chi} h^2)_{min}$ which is obtained
from Eq. (\ref{omega}), by replacing $\widetilde{<\sigma_{ann} v>}$
with its maximal value $(\widetilde{<\sigma_{ann} v>})_{max}$, at
fixed $m_{\chi}$. In the case of Higgs-exchange contributions, this
$(\widetilde{<\sigma_{ann} v>})_{max}$ is obtained by inserting into
Eq. (\ref{ah}) the maximal value of the product $a_1^2 a_3^2 \tan^2
\beta$.  Taking into account Eq. (\ref{ratio}) and that, for $m_A
\simeq $ 90 GeV, the upper bound of $\tan \beta$ is 45 \cite{cdf},
we obtain $(a_1^2 a_3^2 \tan^2 \beta)_{max} \simeq 2.6 \times 10^2$,
and in turn

\begin{eqnarray}
(\Omega_{\chi} h^2)_{min}^{Higgs} &\equiv&  
 \frac{1.5 \cdot 10^{-10}}{\rm GeV^{2}}
 \frac{x_f}{{g_{\star}(x_f)}^{1/2}}
\frac{m_W^2}{m_{\chi}^2} [(2m_{\chi})^2 - m_A^2]^2
\left( \sum_f \bar{m}_f^2  (1 + \epsilon_f)^2 ~ c_f  
~[1-m_f^2/m_\chi^2]^{1/2}\right)^{-1}      \nonumber \\
&\simeq& \frac{ 5 \cdot 10^{-11}}{\rm GeV^{2}} \frac{x_f}{{g_{\star}(x_f)}^{1/2}} 
\frac{m_W^2}{\bar{m}_b^2} \frac{1}{(1 + \epsilon_b)^2 }
 \frac{[(2m_{\chi})^2 - m_A^2]^2}{m_{\chi}^2~[1-m_b^2/m_\chi^2]^{1/2}}.   
\label{omegamin} 
\end{eqnarray}

In the last step of this equation we have retained only the dominant
contribution due to the $b-\bar{b}$ final state.  As far as the value
of ${g_{\star}(x_f)}^{1/2}$ is concerned, we notice that for these
light neutralinos  $x_f \simeq 21-22$, so that neutralinos with masses 
$m_{\chi} \simeq $ 6-7 GeV have
a freeze-out temperature $T_f \sim T_{QCD}$, where $T_{QCD}$ is the
hadron-quark transition temperature of order 300 MeV.  For
definiteness, we describe here the hadron-quark transition by a step
function: if $T_{QCD}$ is set equal to 300 MeV, then for $m_{\chi}
\lsim $ 6 GeV one has ${g_{\star}(x_f)}^{1/2} \simeq 4$, while for
heavier neutralinos ${g_{\star}(x_f)}^{1/2} \simeq 8-9$.

The quantity $(\Omega_{\chi} h^2)_{min}^{Higgs}$ as given by
Eq. (\ref{omegamin}) is plotted in Fig.1 as a function of $m_{\chi}$,
for the value $m_A =$ 90 GeV (current experimental lower bound).  The
solid curve refers to the case $T_{QCD} =$ 300 MeV; dashed and
dot-dashed curves denote two different representative values of
$T_{QCD}$: $T_{QCD} =$ 100 MeV and $T_{QCD} =$ 500 MeV,
respectively. The two horizontal lines denote two representative
values for the upper bound on $\Omega_{CDM} h^2$: $(\Omega_{CDM}
h^2)_{max} =$ 0.3 (short-dashed line) and $(\Omega_{CDM} h^2)_{max} =$
0.131 (long-dashed line).  Fig.1 displays how a lower bound on
$m_{\chi}$ is derived from an upper limit on $\Omega_{CDM} h^2$. In
particular, using $T_{QCD} =$ 300 MeV, one obtains from $(\Omega_{CDM}
h^2)_{max} =$ 0.3 the bound $m_{\chi} \gsim$ 5.2 GeV, a value which
increases to $m_{\chi} \gsim$ 6.2 GeV, when the new value
$(\Omega_{CDM} h^2)_{max} =$ 0.131 is employed.  Also evident is the
variation of the lower bound on $m_{\chi}$, when the value of
$T_{QCD}$ is changed.  Finally, to support the validity of the
analytical approximations employed to derive Eq. (\ref{omegamin}), in
Fig.1 we also display the scatter plot of $\Omega_{\chi} h^2$, when a
numerical scanning of the supersymmetric parameter space is performed.

We recall that the bound on $m_{\chi}$: $m_{\chi} \gsim$ 6.2 GeV was
derived using $m_A =$ 90 GeV (which is the present experimental lower
bound on $m_A$). From our previous formulae one obtains that this
bound on $m_{\chi}$ simply scales with $m_A$ as follows

\begin{equation}
{m_{\chi}~[1-m_b^2/m_\chi^2]^{1/4}} \gsim 5.3 ~ {\rm GeV} \left(
\frac{m_A}{90 \; {\rm GeV}} \right)^2
\label{ma}
\end{equation}

 For the sfermion-exchange contribution, ${\widetilde{\langle
 \sigma_{\rm ann} \; v\rangle}_f}^{sfermion} \simeq
 {\tilde{a}_f}^{sfermion} + \frac{1}{2 x_f} {\tilde{b}_f}^{sfermion}$,
 both S-wave and P-wave contributions have to be taken into account.
 In the regime we consider here, the leading terms due to the
 down-type fermions are

 \begin{equation} {\tilde{a}_f}^{\,sfermion} \simeq \frac{{\pi}
 {\alpha}_{e.m.}^2}{8 \cos^4 \theta_W} a_1^4 c_f \frac{m_f^2 \,
 [1-m_f^2/m_\chi^2]^{1/2}}{{m^4_{\tilde{f},1}} (2 - m_{\tilde{f},1}^2/
 m_{\tilde{f}}^2)^2} \left[(Y_{f,L}^2 + Y_{f,R}^2) + 2 Y_{f,L} Y_{1,R}
 ~ \frac{m_{\chi}}{m_f} ~ \left(1 - \frac{m^2_{\tilde{f},1}}
 {m^2_{\tilde{f}}}\right)\right]^2 \label{af} \end{equation}

 \begin{equation} {\tilde{b}_f}^{\, sfermion} \simeq \frac{{\pi}
 {\alpha}_{e.m.}^2}{4 \cos^4 \theta_W} a_1^4 c_f \frac{m_{\chi}^2 \,
 [1-m_f^2/m_\chi^2]^{1/2}}{{m^4_{\tilde{f},1}} (2 - m_{\tilde{f},1}^2/
 m_{\tilde{f}}^2)^2} \left[2 (Y_{f,L}^4 + Y_{f,R}^4) + 3 Y_{f,L}^2
 Y_{f,R}^2 ~ \left(1 -
 \frac{m_{\tilde{f},1}^2}{m^2_{\tilde{f}}}\right)^2\right].
 \label{bf} \end{equation}

 \noindent Notations are as follows: $Y_{f,L}$ and $Y_{f,R}$ are the
 weak hypercharges for left and right couplings,
 respectively. $m_{\tilde{f}}$ denotes either $m_{\tilde{l}}$ or 
$m_{\tilde{q}}$ depending on the nature of the fermion, 
$m_{\tilde{f},1}$ is the smallest mass eigenvalue for
 the sfermion $\tilde{f}$, once the mass matrix in the weak--interaction basis
 $\tilde{f}_L, \tilde{f}_R$ is diagonalized. 
A maximal mixing between the $\tilde{f}_L, \tilde{f}_R$
 fields has been used in deriving Eqs. (\ref{af})--(\ref{bf}), since
 these equations are meant to provide a conservative lower bound on
 $m_{\chi}$. Notice that for $l = e, \mu$, the $\tilde{f}_L-\tilde{f}_R$ 
 mixing is negligible; this case is simply recovered from
 Eqs. (\ref{af})-(\ref{bf}) by setting $m_{\tilde{f},1} =
 m_{\tilde{f}}$.
 
 With the aid of numerical evaluations, it is found that the leading
 contributions to ${\widetilde{\langle \sigma_{\rm ann} \;
 v\rangle}^{sfermion}} \equiv \sum_{f} {{\widetilde{\langle
 \sigma_{\rm ann} \; v\rangle}_{f}^{sfermion}}}$  are 
provided by the
 term due to the $\tau$ lepton: ${\widetilde{\langle \sigma_{\rm ann}
 \; v\rangle}^{sfermion}} \simeq {\widetilde{\langle \sigma_{\rm ann}
 \; v\rangle}_{\tau}^{sfermion}}$.  This is in turn maximized by

\begin{equation}
({\widetilde{\langle \sigma_{\rm ann} \; v\rangle}^{sfermion}})_{max}
 \simeq \frac{{\pi} {\alpha}_{e.m.}^2}{8 \cos^4 \theta_W} \frac{m_{\chi}^2 \,
 [1-m_\tau^2/m_\chi^2]^{1/2}}{{m_{\tilde{\tau}}}^4} \left[ \left(2 +
 \frac{5}{2} ~ \frac{m_{\tau}}{m_{\chi}}\right)^2 + \frac{23}{2 x_f}
 \right]
\label{sf}
\end{equation}

We denote by $(\Omega_{\chi} h^2)_{min}^{sfermion}$ the value of
$\Omega_{\chi} h^2$ derived from Eq. (\ref{omega}) when
$\widetilde{\langle \sigma_{\rm ann} \; v\rangle}$ is replaced by
$({\widetilde{\langle \sigma_{\rm ann} \;
v\rangle}^{sfermion}})_{max}$ of Eq. (\ref{sf}).  The quantity
$(\Omega_{\chi} h^2)_{min}^{sfermion}$ is plotted in Fig.2, for the
value $m_{\tilde{\tau}} =$ 87 GeV (current experimental lower bound);
also the scatter plot for the quantity $\Omega_\chi h^2$ is
displayed, for $m_A>$300 GeV. 
From Fig.2 one finds for $m_{\chi}$ a lower bound $m_{\chi}
\gsim 14$ GeV for $(\Omega_{CDM} \; h^2)_{max} = 0.3$, and $m_{\chi}
\gsim 22$ GeV for $(\Omega_{CDM} \; h^2)_{max} = 0.131$.  The scaling
of this last bound with the stau mass is approximately given by

\begin{equation}
m_{\chi}~[1-m_{\tau}^2/m_\chi^2]^{1/4} \gsim 
 22 \; {\rm GeV} \; 
\left( \frac{m_{\tilde{\tau}}}{90 \; {\rm GeV}} \right)^2. 
\label{tau}
\end{equation}

In general, one has conservatively to retain as a lower bound to
$m_{\chi}$ the smaller of the two lower limits given separately in
Eq. (\ref{ma}) and in Eq. (\ref{tau}). From these equations one finds
that the lower bound of Eq. (\ref{ma}) is less stringent than the one
of Eq. (\ref{tau}) as long as $m_A \lsim 2 m_{\tilde{\tau}}$. Due to
the present experimental bounds on $m_A$ and $m_{\tilde{\tau}}$ the
lower absolute bound is the one derived from Eq. (\ref{ma}), {\it
i. e.}  $m_{\chi} \gsim$ 6.2 GeV.  
We parenthetically note that the
lower limits $m_{\chi} \gsim (15-18)$ GeV found in
Refs. \cite{boudjema}--\cite{hooper} are due to the assumption that
$m_A$ is very large ($m_A \sim 1$ TeV).

A plot which shows the transition between the lower bounds established
by considering Higgs-exchange contributions and sfermion-exchange
contributions is reported in Fig.3.  As was derived by use of the
approximate expressions in Eqs. (\ref{ma})-(\ref{tau}), the transition
point is at $m_A \sim $ 200-300 GeV.

\section{Detectability of light neutralinos by WIMP direct measurements}

Now we wish to show that a part of the neutralino population of small
$m_{\chi}$ is indeed explorable with experiments of WIMP direct
detection with current sensitivities.  Let us start by giving an
approximate relation between the scalar neutralino-nucleon
cross-section and the neutralino relic abundance, valid when
$\sigma_{\rm scalar}^{(\rm nucleon)}$ is dominated by $h$-exchange in
the $t$-channel and $\sigma_{ann}$ by $A$-exchange in the annihilation
$s$-channel.  From the approximate formulae given in Ref. \cite{light}
for $\sigma_{\rm scalar}^{(\rm nucleon)}$ and $\Omega_{\chi} h^2$ we
obtain

\begin{equation}
(\Omega_{\chi} h^2) \; \sigma_{\rm scalar}^{(\rm nucleon)} \simeq 1.4
\times 10^{-40} \;{\rm cm^2 \; T} \; \left( \frac{m_s ~ \langle N|\bar{s} s |N
\rangle}{200 ~ {\rm MeV}} \right)^2 \frac{\rm GeV^2}{m_{\chi}^2 \,
[1-m_b^2/m_\chi^2]^{1/2}} \; \left( \frac{m_A}{m_h} \right)^4,
\label{prod}
\end{equation}

\noindent
where $m_h$ is the mass of the lightest CP-even neutral Higgs boson and 
$T$ is given by

\begin{equation}
T = \frac{(a_3 ~ \sin \alpha + a_4 ~ \cos \alpha)^2}{(a_4 ~ \cos \beta
- a_3 ~ \sin \beta)^2} \; \frac{(\sin \alpha + \epsilon_s ~ cos
(\alpha - \beta) ~ \sin \beta)^2}{\sin^2 \beta ~ (1 + \epsilon_b)^2}.
\label{t}
\end{equation}

\noindent
In Eqs. (\ref{prod})-(\ref{t}) $\langle N|\bar{s} s |N \rangle$ is the
$s$-quark density matrix element over the nucleonic state and $\alpha$
is the angle which rotates the Higgs fields $H_1^{(0)}$ and
$H_2^{(0)}$ into the mass eigenstates $h$ and $H$. The approximations
employed in deriving Eqs.(\ref{prod})-(\ref{t}) imply $\alpha \sim
\frac{\pi}{2}$, $m_h \sim m_A \sim $ 100 GeV, so that $T$ is of order
one.

Thus, for neutralino configurations with  $m_{\chi} \lsim $ 20 GeV,  
$\sigma_{\rm scalar}^{(\rm nucleon)}$ turns out to be bounded by

\begin{equation}
\sigma_{\rm scalar}^{(\rm nucleon)} \gsim \frac{10^{-40} {\rm cm^2}}
{(\Omega_{CDM} h^2)_{max}} \; \frac{\rm GeV^2}{m_{\chi}^2 \,
[1-m_b^2/m_\chi^2]^{1/2}}.
\label{bound}
\end{equation}

\noindent 
In deriving Eq. (\ref{bound}), we have set $m_s ~ \langle N|\bar{s} s
|N \rangle = $ 200 MeV \cite{size}.

The results of a complete numerical evaluations of the quantity $\xi
\sigma_{\rm scalar}^{(\rm nucleon)}$, where all relevant diagrams for
the neutralino-nucleon scalar cross-section and for the relic
abundance are taken into account, are displayed in Fig.4.  The
peculiar funnel in the scatter plot for $m_{\chi} \lsim $ 20 GeV is due
to the bound of Eq. (\ref{bound}).

As was pointed out in Ref. \cite{light}, the present upper limits to
$\xi \; \sigma_{\rm scalar}^{(\rm nucleon)}$ provided by WIMP direct
detection experiments \cite{morales,dama0,cdms,edel} do not
significantly constrain the supersymmetric configurations for the
light neutralinos displayed in Fig.4. Instead, these configurations
may be relevant for experiments of direct detection with a low energy
threshold and a large exposure.  An experiment with these features is
the DAMA/NaI experiment, whose results, after a 4-years running with a
total exposure of $\simeq$ 58 000 kg $\cdot$ day, show an
annual-modulation effect at a $4\sigma$ C.L. which does not appear to
be related to any possible source of systematics \cite{dama}. The
analysis carried out by the DAMA Collaboration to explain their
modulation effect in terms of a WIMP with coherent elastic scattering
was targeted to a neutralino in the frame of a usual supersymmetric
scheme with gaugino-mass unification at GUT, with a consequent lower
bound on the neutralino mass above 30 GeV.  This interpretation was
proved to be consistent with supersymmetric models with
gaugino-unification at GUT \cite{altri}.

Here we have considered a different supersymmetric scheme which
includes significantly lower neutralino masses; thus, in order to
establish the possible relevance of our low-mass neutralinos for an
annual-modulation effect, we have to proceed to an extension of
previous analyses.  To put our arguments into a quantitative basis, we
evaluate $(\xi \sigma_{\rm scalar}^{(\rm nucleon)})_{min}$, defined as
the minimal value of $\xi \sigma_{\rm scalar}^{(\rm nucleon)}$ which
may produce an annual-modulation effect at $n$ standard deviations in a
detector with a given exposure $S$, and for a given velocity
distribution function $f(\vec{v})$ for relic neutralinos in our
galaxy, {\it i. e.}

\begin{equation}
(\xi \sigma_{\rm scalar}^{(\rm nucleon)})_{min}  = \frac{n^2}{S} 
\frac{I}{(\Delta I)^2}. 
\label{min} 
\end{equation}

\noindent
In Eq. (\ref{min}) $I$ is defined as the ratio of the expected direct
detection rate, integrated over an energy range ($E_1, E_2$), to the
neutralino-nucleon scalar cross-section $\xi \sigma_{\rm scalar}^{(\rm
nucleon)}$:

\begin{equation}
I = \frac{1}{\xi \sigma^{\rm (nucleon)}_{\rm scalar}}
\int_{E_1}^{E_2}\frac{d R(E)}{d E} dE.
\end{equation}

\noindent
For a monoatomic material of nuclear mass number $A$, one has 

 \begin{equation}
I = N_T \rho_0 \frac{m_N}{2 m^3_{\chi}}\left
(1+\frac{m_{\chi}}{m_p}\right )^2 A^2 \int_{E_1}^{E_2} dE
F^2(E)\int_{v_{\rm \min}(E)}^{\infty}d \vec{v}\frac{f(\vec{v})}{|\vec{v}|}, 
\label{I}
 \end{equation}

\noindent
where $N_T$ is the number of the target nuclei per unit of mass, $m_N$
is the nuclear mass, $F(E)$ the nuclear form factor and $v_{min}$ is
the minimal value of the neutralino velocity to produce an event above
the detection threshold.  Generalization of Eq. (\ref{I}) to non-monoatomic 
materials is straightforward.  In Eq. (\ref{min}) $\Delta I$ is
defined as $\Delta I \equiv (I({\rm June}) - I({\rm December}))/2$.

We emphasize that, by definition, the condition $\xi\sigma_{\rm
scalar}^{(\rm nucleon)} \geq (\xi \sigma_{\rm scalar}^{(\rm
nucleon)})_{min}$ only establishes a minimal requirement for a
neutralino of a given mass $m_{\chi}$ to be detected in a given
experiment.  It is obvious that the real detectability of the signal
depends on the actual amount of the experimental background. For
example, an equal amount of signal and background would double the
value of $(\xi \sigma_{\rm scalar}^{(\rm nucleon)})_{min}$ as given in
Eq. (\ref{min}). Estimates of backgrounds are omitted here.

To establish whether our population of light neutralinos may be
relevant for the effect measured by the DAMA experiment, we have
evaluated $(\xi \sigma_{\rm scalar}^{(\rm nucleon)})_{min}$ as a
function of $m_{\chi}$; the energy range of integration employed here
is: $E_1$ = 2 keV, $E_2$ = 3 keV, and $n$ is set equal to 4. The
nuclear form factors for $Na$ and $I$ are modelled in the Helm form
\cite{helm}, with the values of parameters given in Ref. \cite{exp}.
In Fig.4 we give the curves of $(\xi \sigma_{\rm scalar}^{(\rm
nucleon)})_{min}$ versus $m_{\chi}$ for a sample of different galactic
distribution functions (DF), taken among those analysed in
Ref. \cite{bcfs}. The intermediate curve refers to an isothermal DF
with $v_0 = 220$ km $\cdot$ s$^{-1}$ and $\rho_0 = 0.3$ GeV $\cdot$
cm$^{-3}$ ($v_0$ is the local rotational velocity). The upper curve
refers to a spherical Evans' power-law DF (denoted as A3 in
Ref. \cite{bcfs}) with $v_0 = 170$ km $\cdot$ s$^{-1}$ and $\rho_0 =
0.17$ GeV $\cdot$ cm$^{-3}$; the lower curve refers to an
axially-symmetric Evans' logarithmic DF with maximal flatness (denoted
as C2 in Ref. \cite{bcfs}) with $v_0 = 270$ km $\cdot$ s$^{-1}$ and
$\rho_0 = 1.7$ GeV $\cdot$ cm$^{-3}$.

From the results displayed in Fig.4 one sees that indeed a part of our
population of relic neutralinos of small $m_{\chi}$ may be relevant
for the annual--modulation effect discussed above, especially in case
of DF's which entail higher velocities and /or higher densities as
compared to the standard isothermal distribution. This is a new
interesting option which adds to the other, still valid, possibility
which we discussed in the papers of Ref. \cite{altri}, on relic
neutralinos with masses above 50 GeV.

\section{Conclusions}

We have considered phenomenological properties of relic neutralinos in
a range of low masses ($m_{\chi} \lsim$ 45 GeV), which is allowed in a
MSSM model without gaugino-mass unification at a grand unification
 scale. We have numerically evaluated the relevant relic
abundance and the neutralino-nucleon cross-section, and discussed our
numerical results in terms of analytic formulae, which display the
connections among cosmological properties and particle-physics
parameters.
   
 Using the latest determinations of $\Omega_{CDM} h^2$ 
and assuming R-parity conservation, we have shown
 that in a MSSM model without gaugino-mass unification the lower bound
 on the neutralino mass is $m_{\chi} \gsim 6$ GeV. We have shown
 analytically how this bound is linked to the cosmological upper bound
 on $\Omega_{CDM} h^2$ and to the lower limits on masses of Higgs
 bosons and sfermions.
   
The implications of light relic neutralinos with masses $ \lsim $ 45
GeV for WIMP direct searches have been analysed.  It is found that
these neutralinos are actually relevant for WIMP direct detection
experiments of low energy threshold and large exposure.  The present
results extend to small masses our previous analyses about the effects
of relic neutralinos with masses above 50 GeV \cite{altri}.

\acknowledgements We acknowledge Research Grants funded jointly by
Ministero dell'Istruzione, dell'Universit\`a e della Ricerca (MIUR),
by Universit\`a di Torino and by Istituto Nazionale di Fisica Nucleare
within the {\sl Astroparticle Physics Project}.

\newpage

\renewcommand{\thefigure}{\arabic{figure}}
\begin{figure} \centering
\includegraphics[width=1.0\columnwidth]{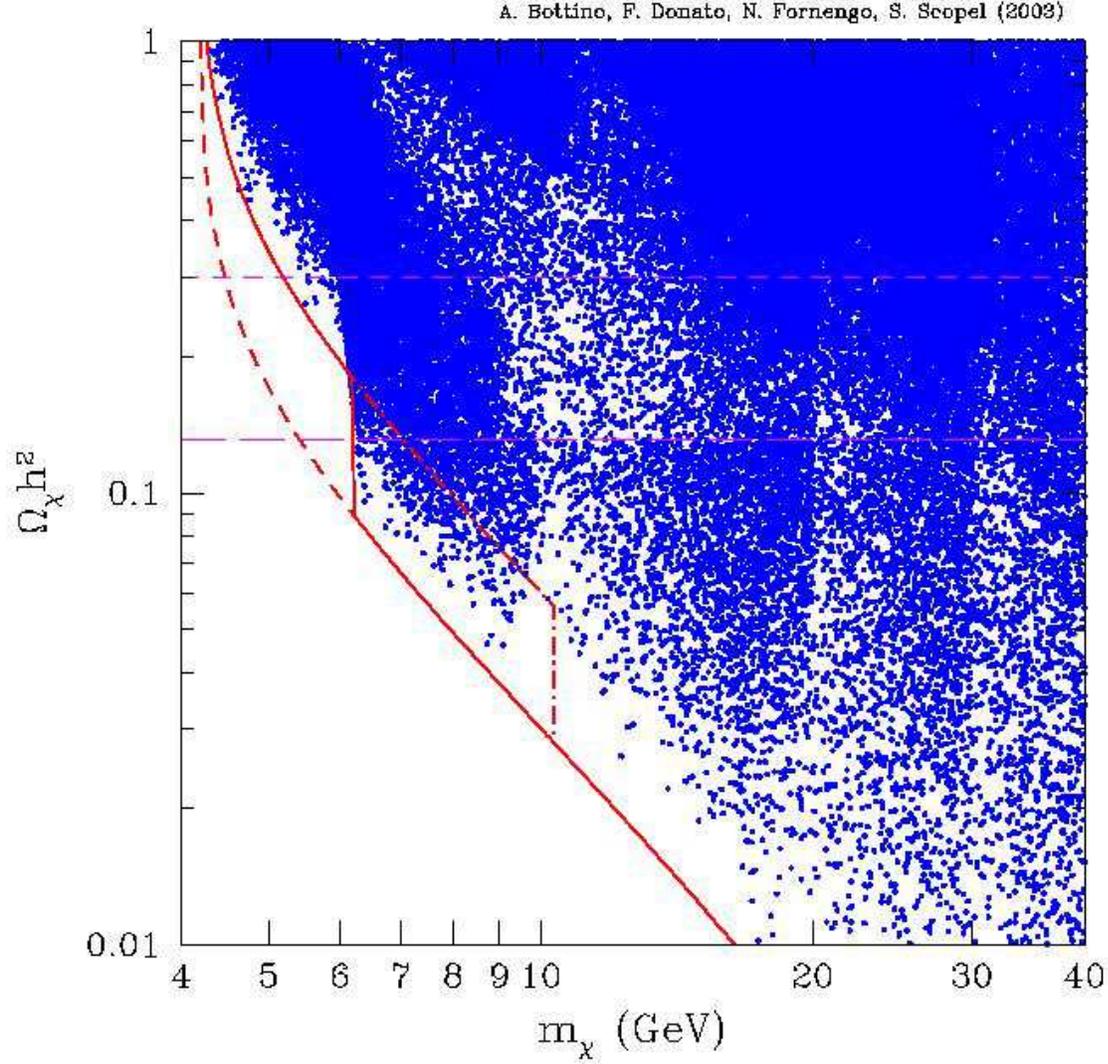}
\caption{\label{fig:1}
Neutralino relic abundance $\Omega_{\chi}h^2$ as a function of the 
 mass $m_\chi$.
 The solid curve denotes $(\Omega_{\chi}
h^2)_{min}^{Higgs}$ given by Eq. (\ref{omegamin}) for $T_{QCD} =$ 300
MeV.  Dashed and dot-dashed curves refer to the representative values
$T_{QCD} =$ 100 MeV, $T_{QCD} =$ 500 MeV, respectively.  The two
horizontal lines denote two representative values of $\Omega_{CDM}
h^2$: $\Omega_{CDM} h^2 =$ 0.3 (short-dashed line) and $\Omega_{CDM}
h^2 =$ 0.131 (long-dashed line). The scatter plot is obtained by a
full scanning of the supersymmetric parameter space.  }
\end{figure}

\begin{figure} \centering
\includegraphics[width=1.0\columnwidth]{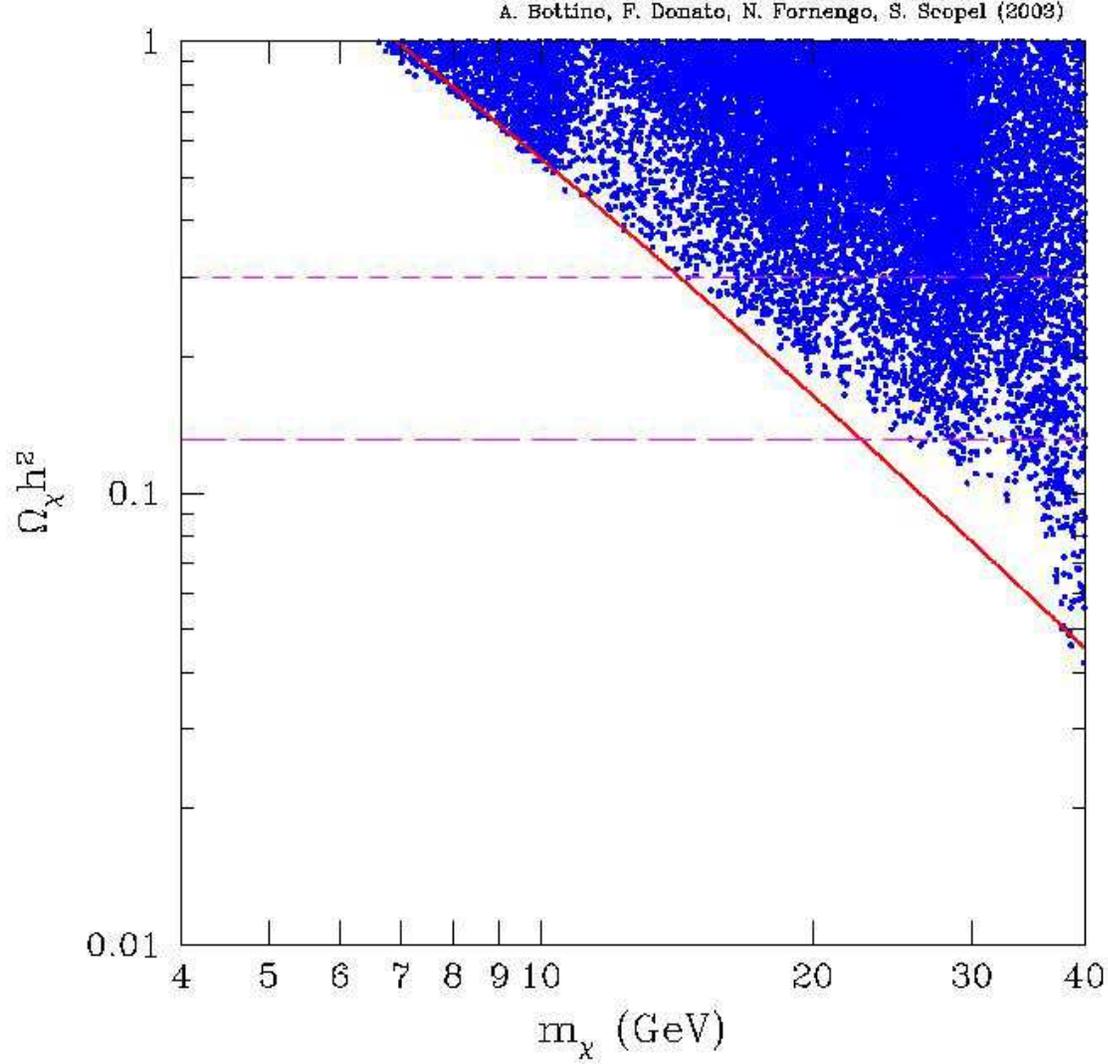}
\caption{\label{fig:2} 
Neutralino relic abundance $\Omega_{\chi}h^2$ as a function of the 
 mass $m_\chi$.
The solid curve denotes $(\Omega_{\chi}
h^2)_{min}^{sfermion}$ derived from Eq. (\ref{omega}) when
$\widetilde{\langle \sigma_{\rm ann} \; v\rangle}$ is given by the
maximal value of ${\widetilde{\langle \sigma_{\rm ann} \;
v\rangle}^{sfermion}}$ (see Eq. (\ref{sf})).  $T_{QCD}$ is set equal
to 300 MeV. The two horizontal lines denote two representative values
of $\Omega_{CDM} h^2$: $\Omega_{CDM} h^2 =$ 0.3 (short-dashed line)
and $\Omega_{CDM} h^2 =$ 0.131 (long-dashed line).  The scatter plot
is obtained by a full scanning of the supersymmetric parameter space
with $m_A>$ 300 GeV.
}
\end{figure}

\begin{figure} \centering
\includegraphics[width=1.0\columnwidth]{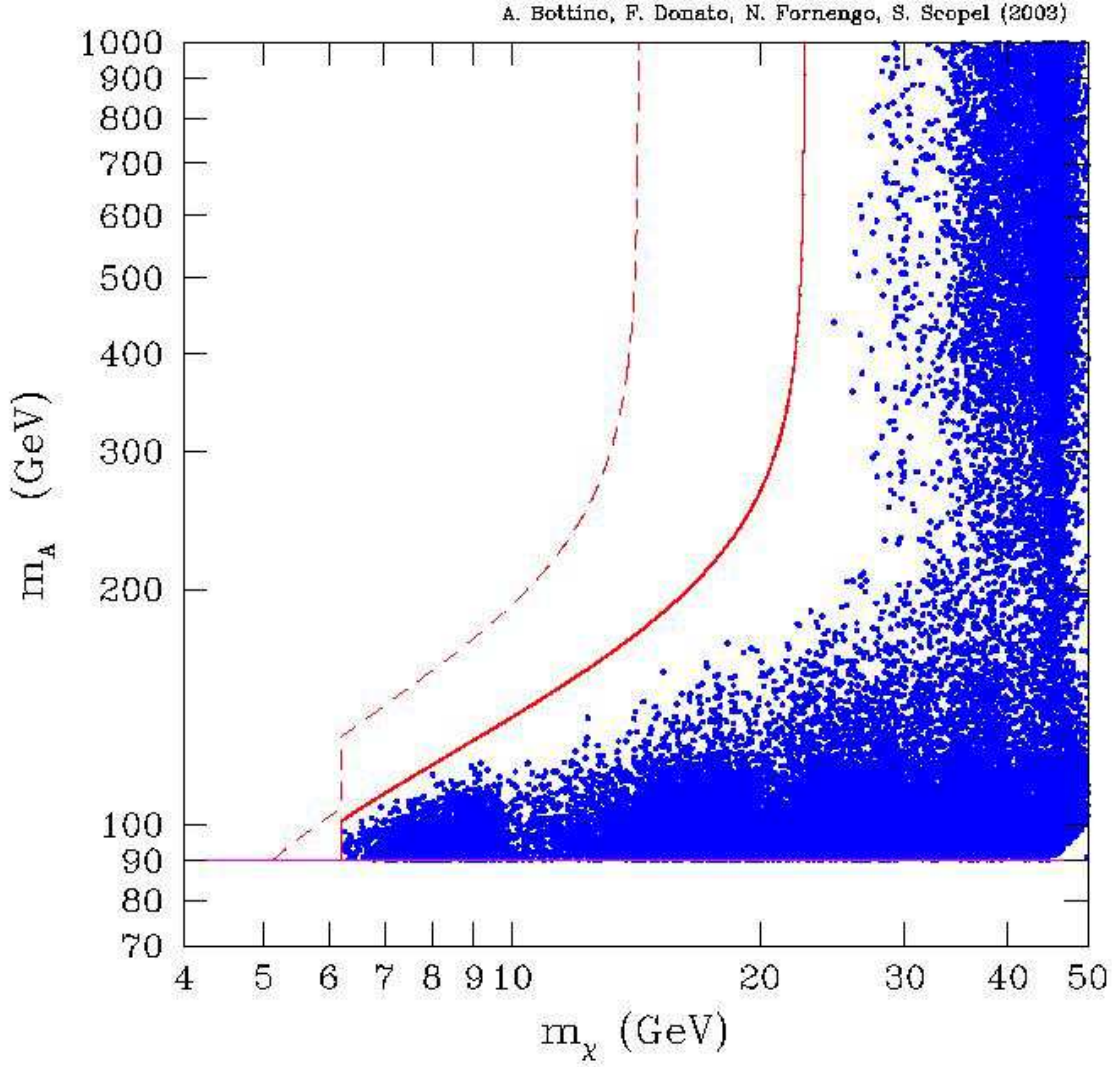}
\caption{\label{fig:3} Dashed and solid curves give the variation of
the lower bound on $m_{\chi}$ as a function of $m_A$ for
$(\Omega_{CDM} h^2)_{max} =$ 0.3 and for $(\Omega_{CDM} h^2)_{max} =$
0.131, respectively. For each value of $(\Omega_{CDM} h^2)_{max}$ the
region on the left of the relevant curve is forbidden.  The two lines
are derived from the analytical expressions obtained in the text for
the contributions to $\widetilde{\langle \sigma_{\rm ann} \;
v\rangle}$ due to Higgs-exchange and sfermion-exchange. Also displayed
is the scatter plot of a full numerical scanning.  }
\end{figure}

\begin{figure} \centering
\includegraphics[width=1.0\columnwidth]{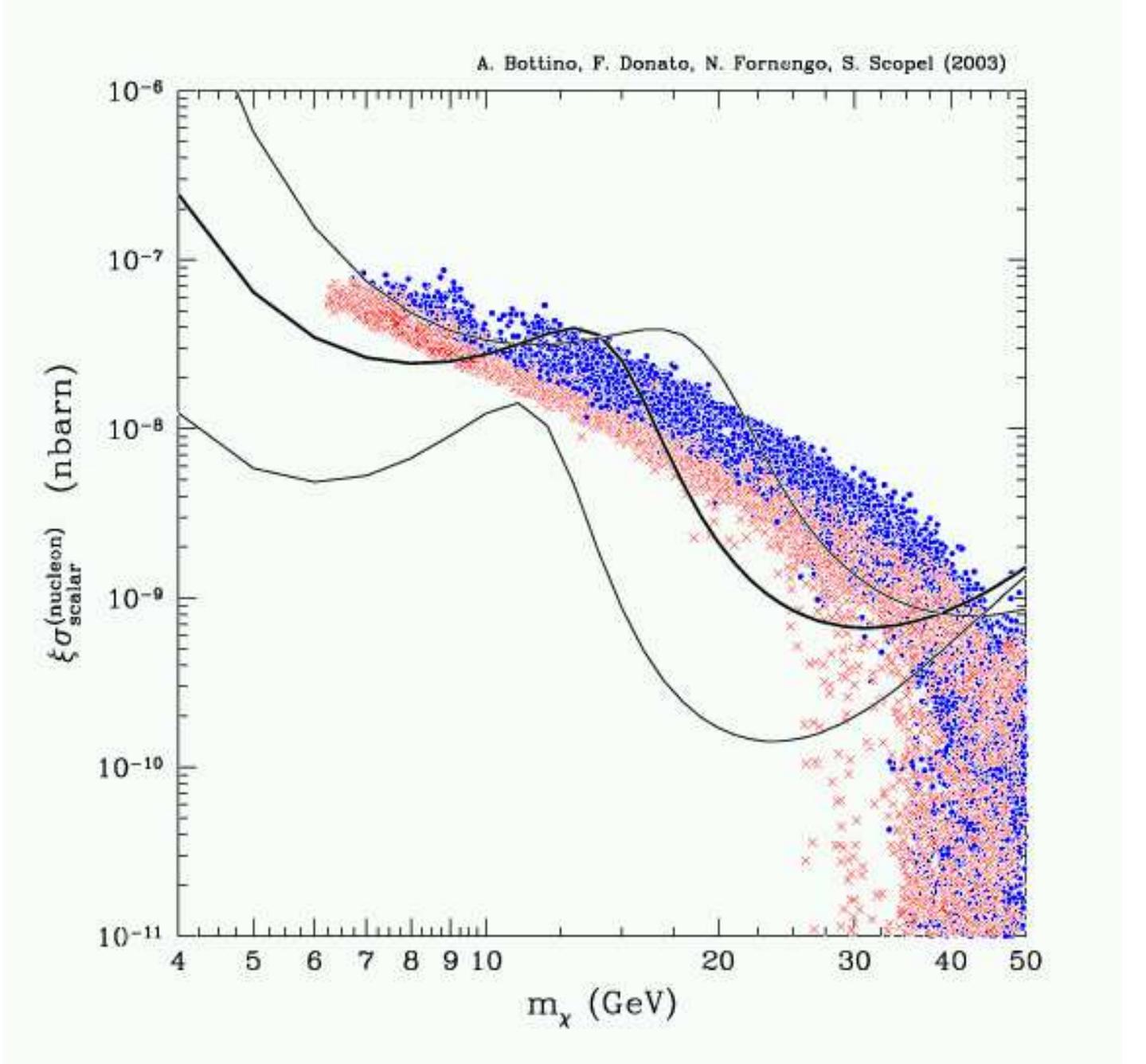}
\caption{\label{fig:4} 
Scatter plot of $\xi \sigma_{\rm scalar}^{(\rm
nucleon)}$ versus $m_{\chi}$.  Crosses (red) and dots (blue) denote
neutralino configurations with $\Omega_{\chi} h^2 > (\Omega_{CDM}
h^2)_{min}$ and $\Omega_{\chi} h^2 < (\Omega_{CDM} h^2)_{min}$,
respectively.  The curves give the sensitivity line, $(\xi \sigma_{\rm
scalar}^{(\rm nucleon)})_{min}$ versus $m_{\chi}$, for a NaI detector,
whose features are discussed in the text.  The intermediate curve
refers to an isothermal DF with $v_0 = 220$ km $\cdot$ s$^{-1}$ and
$\rho_0 = 0.3 ~ {\rm GeV \cdot cm}^{-3}$. 
The upper curve refers to a spherical Evans' power-law DF
(denoted as A3 in Ref. \cite{bcfs}) with $v_0 = 170$ km $\cdot$
s$^{-1}$ and $\rho_0 = 0.17 ~ {\rm GeV \cdot cm}^{-3}$; the lower curve
refers to an axially-symmetric Evans' logarithmic DF with maximal
flatness (denoted as C2 in Ref. \cite{bcfs}) with $v_0 = 270$ km
$\cdot$ s$^{-1}$ and $\rho_0 = 1.7 ~ {\rm GeV \cdot cm}^{-3}$.  }
\end{figure}

\end{document}